\documentstyle[pre,aps,epsf,amsfonts,floats]{revtex}

\begin{document} 

\flushbottom

\draft
\twocolumn[\hsize\textwidth\columnwidth\hsize\csname @twocolumnfalse\endcsname

\title{Crossover phenomenon in self-organized critical sandpile models}

\author{S. L\"ubeck\cite{SvenEmail} }

\address{
Theoretische Tieftemperaturphysik, 
Gerhard-Mercator-Universit\"at Duisburg,\\ 
Lotharstr. 1, 47048 Duisburg, Germany \\}

\date{Received 10.\,April 2000}

\maketitle

\begin{abstract}
We consider a stochastic sandpile where the sand-grains 
of unstable sites are randomly distributed to the 
nearest neighbors.
Increasing the value of the threshold condition 
the stochastic character of the distribution
is lost and a crossover to the scaling behavior
of a different sandpile model takes place where the sand-grains
are equally transferred to the nearest neighbors.
The crossover behavior is numerically analyzed in detail,
especially we consider the exponents
which determine the scaling behavior.
\end{abstract}


]  

\setcounter{page}{1}
\markright{\rm accepted for publication in Physical Review E {\bf 62}, (2000)}
\thispagestyle{myheadings}
\pagestyle{myheadings}

\section{Introduction}

Crossovers between different universality
classes are well known from equilibrium
phase transitions.
Similar crossover phenomena are also known from
non-equilibrium systems which exhibit
self-organized criticality~\cite{BAK_1}.
For instance  the forest-fire model of
Drossel and Schwabl~\cite{DROSSEL_3} displays 
a crossover to a percolation like scaling behavior
if one introduces an immunity parameter
which prevents trees from burning~\cite{DROSSEL_4}.
Another example is the directed Abelian sandpile
model of Dhar and Ramaswamy~\cite{DHAR_1}.
Introducing a second stochastic toppling condition
the system changes to a directed percolation like 
scaling behavior~\cite{TADIC_2}.

In this paper we investigate a crossover
between the stochastic sandpile model introduced
by Manna~\cite{MANNA_2} and the Zhang sandpile model~\cite{ZHANG_1}.
Following Ben-Hur and Biham~\cite{BENHUR_1}
the crossover connects the different universality
classes of undirected sandpile models (the energy or sand-grain
transfer of the nearest neighbors is isotropic, e.g.~the
Zhang model) 
and undirected on average sandpile models (the energy transfer is
isotropic on average only, e.g.~the Manna model).
Considering the Manna model the crossover
takes place just by increasing the threshold 
value which determines the dynamics, i.e.,
no additional parameter has to be introduced.

In the next section we briefly remind the 
distinctive characteristics of the Zhang model.
These characteristics will allow us in the 
following to identify the typical Zhang scaling behavior.
Then we describe in section~\ref{manna}
the Manna model for different values of the
threshold condition.
The crossover behavior between both models
is investigated in section~\ref{crossover}.
A summary closes the paper.

\section{The Zhang model}
\label{zhang}

Consider the Zhang model~\cite{ZHANG_1}
on a two dimensional square lattice of linear size $L$.
A continuous value $E_{\underline r}\ge0$ representing the energy 
is associated to each lattice site $\underline r$.
A configuration $\{E_{\underline r}\}$ is stable 
if $E_{\underline r}<E_{\rm c}$
for all lattice sites $\underline r$.
For the sake of simplicity we choose in all 
simulations $E_{\rm c}=1$.
A quantum of energy $\delta$ is added to a randomly chosen lattice 
site $\underline r$, i.e.,
\begin{equation}
E_{\underline r}\;\to\;E_{\underline r}\,+\,\delta.
\label{eq:perturbation_zhang}
\end{equation}
We consider in this work especially the slow driving 
limit $\delta \ll E_{\rm c}$.
For $\delta \to 0$ all lattice sites grow parallel
and the Zhang model corresponds to the conservative limit of 
the ``spring block'' model of Christensen and 
Olami~\cite{CHRIS_1}.

In the case that due to the perturbation a site exceeds the
critical value $E_{\rm c}$ an activation event will occur, the
unstable site relaxes to zero, and the energy is added
to the nearest neighbors, i.e.,
\begin{equation}
E_{\underline r}\;\to\;0,
\label{eq:relaxation_1_zhang}
\end{equation}
\begin{equation}
E_{{\underline r},NN}\;\to\;E_{{\underline r},NN}\;
+\;\frac{E_{{\underline r}}}{4}.
\label{eq:relaxation_2_zhang}
\end{equation}
The transferred energy may activate the neighboring
sites and thus an avalanche of relaxation events may take place.
Energy may leave the system only at the boundary.
Since the energy of unstable sites is equally 
transfered to the nearest neighbors it was argued
that the Zhang model belongs to the universality class of 
undirected sandpile models~\cite{BENHUR_1}.
It was expected that the Zhang
model and the well-known Bak-Tang-Wiesenfeld (BTW) 
model~\cite{BAK_1} belong to the same universality 
class.
But the scaling behavior of the BTW avalanches
is complex and is not understood.
Although most authors agree upon a breakdown of
simple scaling the interpretation of the numerically
obtained data is still controversial among the 
different groups~(see for instance 
\cite{DEMENECH_1,TEBALDI_1,CHESSA_2,LUEB_2}).
However, we use in the following the classification 
ansatz of~\cite{BENHUR_1} and denote the universality
class of the Zhang model as the class of undirected
sandpile models.

The avalanches are characterized by several physical
properties like the size $s$~(number of relaxation events), 
the area $a$~(number of distinct toppled sites), the
time $t$~(number of parallel updates until the configuration
is stable), the radius $r$ (radius of gyration), etc.
In the critical steady state the corresponding probability
distributions should obey power-law behavior~\cite{BAK_1}
\begin{equation}
P_x (x) \; \sim \; x^{-\tau_x}
\label{eq:power_law}
\end{equation}
characterized by the avalanche exponents $\tau_x$ with 
$x\in\{s,a,t,r\}$.

The Zhang model was intensively investigated in the
last years (see for 
instance~\cite{ZHANG_1,PIETRO_1,JANOSI_1,DIAZ_2,LUEB_3,GIACOMETTI_1}).
A characteristic property of the Zhang model is
the concentration of the steady state energy distribution $p(E)$ 
around~$z$ distinct peaks, where $z$ is the number of
nearest neighbors of the lattice~\cite{ZHANG_1,PIETRO_1,JANOSI_1,LUEB_3}.
The peaks are located at multiples of $(z+1)/z^2$ 
and the height of the peaks diverges in the thermodynamic
limit $L\to \infty$ (see~\cite{LUEB_3} and references therein).
In the case of an infinite lattice the energy distribution
is given by 
\begin{equation}
p(E)\,=\,\sum_{i=0}^{z-1} \,f_i \;\delta(E-E_i),
\label{eq:p_E_infty}
\end{equation} 
where $f_i$ denotes the statistical weight and $E_i$ denotes 
the position of each peak.
It was found numerically that the statistical weights
are independent of the input energy~$\delta$~\cite{LUEB_3}.
Thus the statistical weights can be regarded as another
fingerprint of the Zhang model.

Analyzing the numerically obtained avalanche
distributions~[Eq.~(\ref{eq:power_law})]
it was observed that the avalanche exponents of the 
Zhang model exhibit finite-size corrections
according to~\cite{LUEB_3,GIACOMETTI_1}
\begin{equation}
\tau_x(L) \; = \; \tau_x \, - \, \frac{\rm const_x}{L}.
\label{eq:zhang_tau_L}
\end{equation}
In this case the values of the infinite 
lattice~$\tau_{x}$ are obtained by
an extrapolation to $L\to \infty$.
More than the explicit values of the exponents
this characteristic system size dependence 
allows us in the following to identify the Zhang like 
scaling behavior.

\section{The Manna model}
\label{manna}

A stochastic sandpile model in which integer
values represent local energies (or number of sand-grains) 
was introduced
by Manna~\cite{MANNA_2}.
Here, unstable sites relax to zero
if $E_{\bf r}\ge E_{\rm c}$
and the removed energy is randomly distributed
to the nearest neighbors in the way that one 
chooses randomly for each energy unit (one sand-grain) one neighbor.
Thus the Manna model is characterized by
a stochastic energy transfer and according
to~\cite{BENHUR_1} it belongs to the so-called 
universality class of undirected on average sandpile 
models.

Due to the reduction of the energy of unstable
sites to zero both the Zhang and the Manna model
are non-Abelian models~\cite{DHAR_2}, i.e., 
the stable energy configurations depend on the
sequence in which unstable sites are toppled.
Recently Dhar introduced an Abelian version of the 
two-dimensional Manna model where the energy of critical sites
is not reduced to zero but $E_{i,j} \to E_{i,j} - 2$. 
The energy $\Delta E =2$ is then equally distributed with 
probability $1/2$ to the sites ($i\pm1,j$) or otherwise to the 
sites ($i,j\pm1$)~\cite{DHAR_7}.
In this case it is possible to extend an operator
algebra, which was successfully applied in studying the 
Bak-Tang-Wiesenfeld model~\cite{DHAR_2}, to this modified Manna model.

In contrast to this analytically tractable Abelian
Manna model we consider in this paper the
original Manna model ($E_{\underline r}\to 0$).
Numerous numerical analysis of the Manna model 
were performed for $E_{\rm c}=2$ and the values
of the exponents are known within some error-bars
(see for instance~\cite{MANNA_2,BENHUR_1,CHESSA_2,LUEB_9} 
and references therein).
Usually one assumes that the value of the critical
energy $E_{\rm c}$ has no influence on the
scaling behavior of the model~(see for 
instance~\cite{BENHUR_1,LUEB_9})
and indeed the analysis of the avalanche probability
distributions [Eq.~(\ref{eq:power_law})] 
for $E_{\rm c}\in \{2,3,5,10\}$
reveals that the exponents are independent 
of~$E_{\rm c}$ (see Fig.~1 in ref.~\cite{LUEB_9}).
But as we will see a crossover to
a different universality takes place for
sufficiently large values of the critical energy.

Consider the Manna model on a lattice 
with~$z$ nearest neighbors.
Assume that a given site~$\underline r$ exceeds
the critical value, i.e., $E_{\underline r}\ge E_{\rm c}$.
Thus a toppling process takes place and the
unstable energy~$E_{\underline r}$ is randomly distributed
to the nearest neighbors.
A given neighbor of~$\underline r$ get $n$ energy units
and the corresponding probability distribution is
\begin{equation}
P(n, E_{\underline r}) \; = \;
\,{E_{\underline r} \choose n}\; p^{n} \, q^{E_{\underline r}}
\label{eq:prob_energy}
\end{equation}
with $p=1/z$ and $q=1-p$.
The corresponding expectation value of the energy transfer is 
$\mu =p E_{\underline r}$ and its variance 
$\sigma = \sqrt{p q E_{\underline r}\,}$.
One can therefore distinguish between two regimes:
for small values of $E_{\underline r}$ the expectation 
value is of the same order as the variance ($\mu\approx \sigma$),
i.e., the energy transfer to the nearest neighbors
is characterized by strong fluctuations.
With increasing unstable energies the fluctuations 
decreases and they can be neglected
if $\mu \gg \sigma$ for large values of $E_{\rm r}$.

Since the value of the unstable energy~$E_{\underline r}$ 
is of the same order as the critical energy~$E_{\rm c}$
we introduce the crossover parameter  
$K=\sqrt{(z-1)/E_{\rm c}}$.
As long as the fluctuations are relevant $K\approx 1$
we expect that the scaling behavior
agrees with that of the Manna model for small $E_{\rm c}$.
Increasing the threshold value $E_{\rm c}$ the
fluctuations become irrelevant if $K\ll 1$.
In this case the energy is nearly equally distributed
to the nearest neighbors and the model corresponds
to the Zhang model in the limit $\delta\ll E_{\rm c}$.
Thus we expect that a crossover from the Manna to the Zhang 
scaling behavior takes place if one increases
the value of the critical energy.
The details of this crossover are investigated in the
next section.

\section{Crossover behavior}
\label{crossover}

\subsection{Energy distribution and energy correlations}

In this subsection we investigate the static 
properties of the Manna model for various
values of the critical energy~$E_{\rm c}$.
We show that the crossover affects the distribution
of the energies, its scaling behavior as well
as the spatial correlations of the energies.

\begin{figure}[t]
 \epsfxsize=8.0cm
 \epsffile{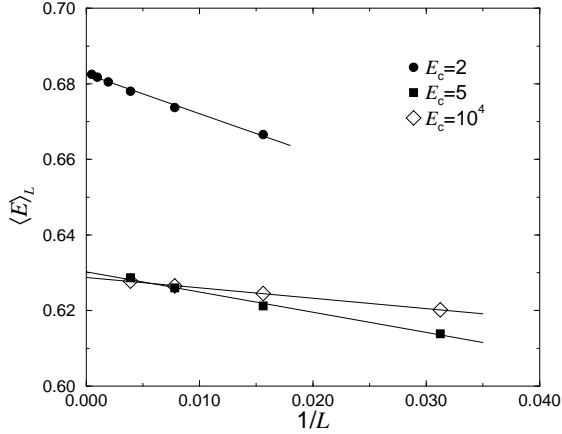}
 \caption{The average energy in the steady state 
          $\langle E \rangle_L$ as a function of the inverse
	  system size~$L$ for different values of~$E_{\rm c}$.
	  The extrapolation to the vertical axis yields
	  the value of the average energy for $L\to \infty$.
 \label{fig:E_aver_01}} 
\end{figure}

We measured the average energy in the steady state
\begin{equation}
\langle E \rangle_L \; = \;
\left \langle L^{-2} \sum_{i,j} \, E_{i,j} 
\right \rangle
\label{eq:aver_energy}
\end{equation}
for various values of $L$ and $E_{\rm c}$.
Figure~\ref{fig:E_aver_01} shows the system size 
dependence of the average energy $\langle E \rangle_L$
for three different values of~$E_{\rm c}$.
For all considered values of~$E_{\rm c}$ we observed
that the system size dependence of the average
energy is given by 
\begin{equation}
\langle E \rangle_L \; = \; \langle E \rangle_{L \to \infty}
\, - \, {\rm const} / L.
\label{eq:aver_energy_L}
\end{equation}
This behavior was already found by Manna for 
$E_{\rm c}=2$ \cite{MANNA_2} and it is also
known from the Bak-Tang-Wiesenfeld 
model~\cite{GRASS_1}.
The origin of this system size dependence are boundary effects.
The average energy on the boundary is
smaller than the energy in the bulk
and the relative number of boundary
sites scales as $L^{-1}$.

\begin{figure}[b]
 \epsfxsize=8.0cm
 \epsffile{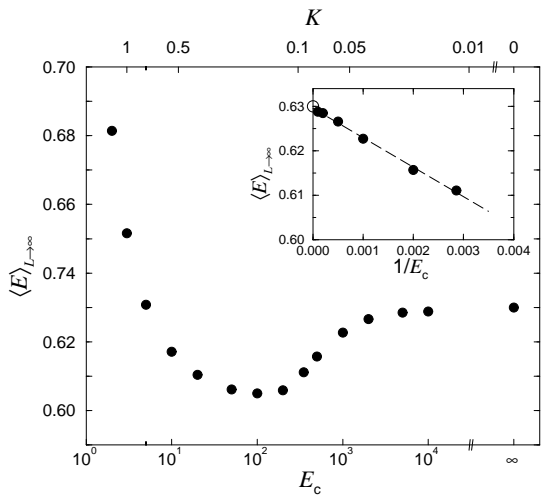}
 \caption{The average energy in the steady state 
          $\langle E \rangle_{L\to \infty}$ as a function of 
	  the critical energy~$E_{\rm c}$.
	  The inset displays the average energy vs.~$1/E_{\rm c}$
	  for $E_{\rm c} \protect\ge 350$.
	  The corresponding value of the pure Zhang model
	  is marked as $E_{\rm c}=\infty$ and as an open circle
          (inset), respectively.
 \label{fig:E_aver_02}}
\end{figure}

The extrapolation to $L\to \infty$ yields the 
average energy of the infinite system size and the obtained
values are shown in Fig.~\ref{fig:E_aver_02} as
a function of $E_{\rm c}$.
Increasing the critical energy from $E_{\rm c}=2$
the average energy $\langle E \rangle_{L\to \infty}$ 
decreases and reaches a minimum for 
$E_{\rm c}\approx 100$.
Here the behavior changes and the average energy increases
with the critical energy.
For large values of $E_{\rm c}$ the average energy 
saturates in the vicinity of the value of the Zhang
model.
A detail analysis suggests that the dependence of
average energy on $E_{\rm c}$ is given by
\begin{equation}
\langle E \rangle_{L \to \infty} \; = \; 
\langle E \rangle_{\rm Zhang}  
\, - \, {\rm const} / E_{\rm c}
\label{eq:aver_energy_Ec}
\end{equation}
for $E_{\rm c} \ge 350$ (see inset of Fig.~\ref{fig:E_aver_02}).
Since finite values of the critical energy results
in finite fluctuations we get 
from Eq.~(\ref{eq:aver_energy_Ec}) that the average
energy is affected by these fluctuations even  
for very large values of $E_{\rm c}$.

\begin{figure}[t]
 \epsfxsize=8.0cm
 \epsffile{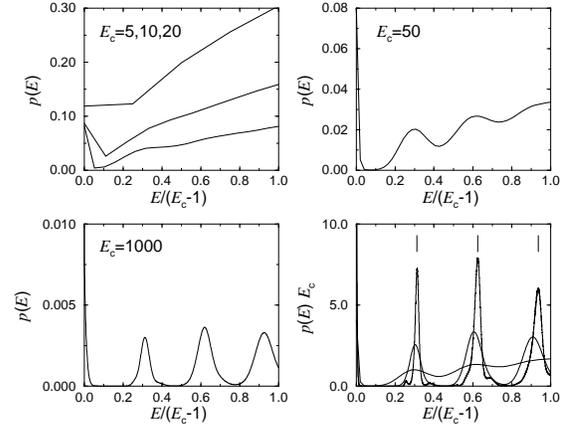}
 \caption{The energy probability distribution $p(E)$ for $L=128$
	  and various values of $E_{\rm c}$.
	  To guide the eye solid lines are plotted instead of symbols.
	  Normalizing $p(E)$ to $E_{\rm c}$ allows to 
	  illustrate how the peak structure of the distribution
	  appears with increasing critical energy.
          This is shown in the lower right figure for 
	  $E_{\rm c}=50, 500, 10000$. 
	  The vertical solid lines correspond to the 
	  position of the peaks of the Zhang model
	  $E_i=5/16 \, i$ with $i=0, 1, 2, 3$~\protect\cite{LUEB_3}.
 \label{fig:p_h_01}} 
\end{figure}

\begin{figure}[b]
 \epsfxsize=8.0cm
 \epsffile{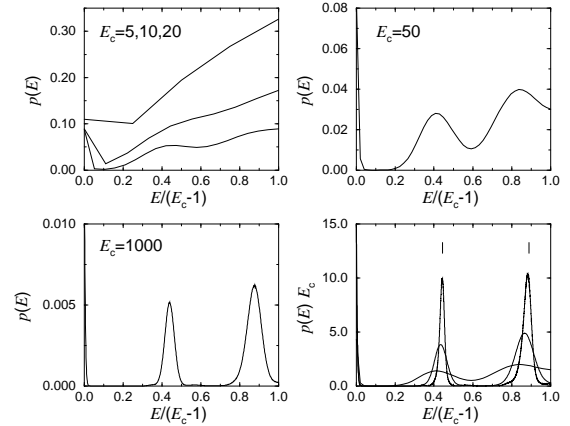}
 \caption{Analogous to Fig.~\ref{fig:p_h_01} but for a
	  honeycomb lattice.
	  The vertical solid lines correspond to the 
	  position of the peaks of the Zhang model
	  $E_i=4/9 \, i$ with $i=0, 1, 2$~\protect\cite{LUEB_3}.
 \label{fig:p_h_01_hc}} 
\end{figure}

After the average value of the energy 
we consider now the steady state 
distribution $p(E)$ of the energies.
In Fig.~\ref{fig:p_h_01} we plot $p(E)$ for various
values of~$E_{\rm c}$.
With increasing critical energy a peak structure appears
with four distinct peaks located at $E_i= 5/16\, i$ 
with $i=0,1,2,3$.
Thus the positions of the peaks agree with 
the corresponding values of the Zhang 
model~(see~\cite{LUEB_3} and reference therein).

A similar analysis on a honeycomb
lattice is shown in Fig.~\ref{fig:p_h_01_hc}.
Here the probability distribution is characterized
by three peaks ($z=3$) located at multiples
of $E=4/9$.
Again these values are in agreement with those 
of the Zhang model.

\begin{figure}[t]
 \epsfxsize=8.0cm
 \epsffile{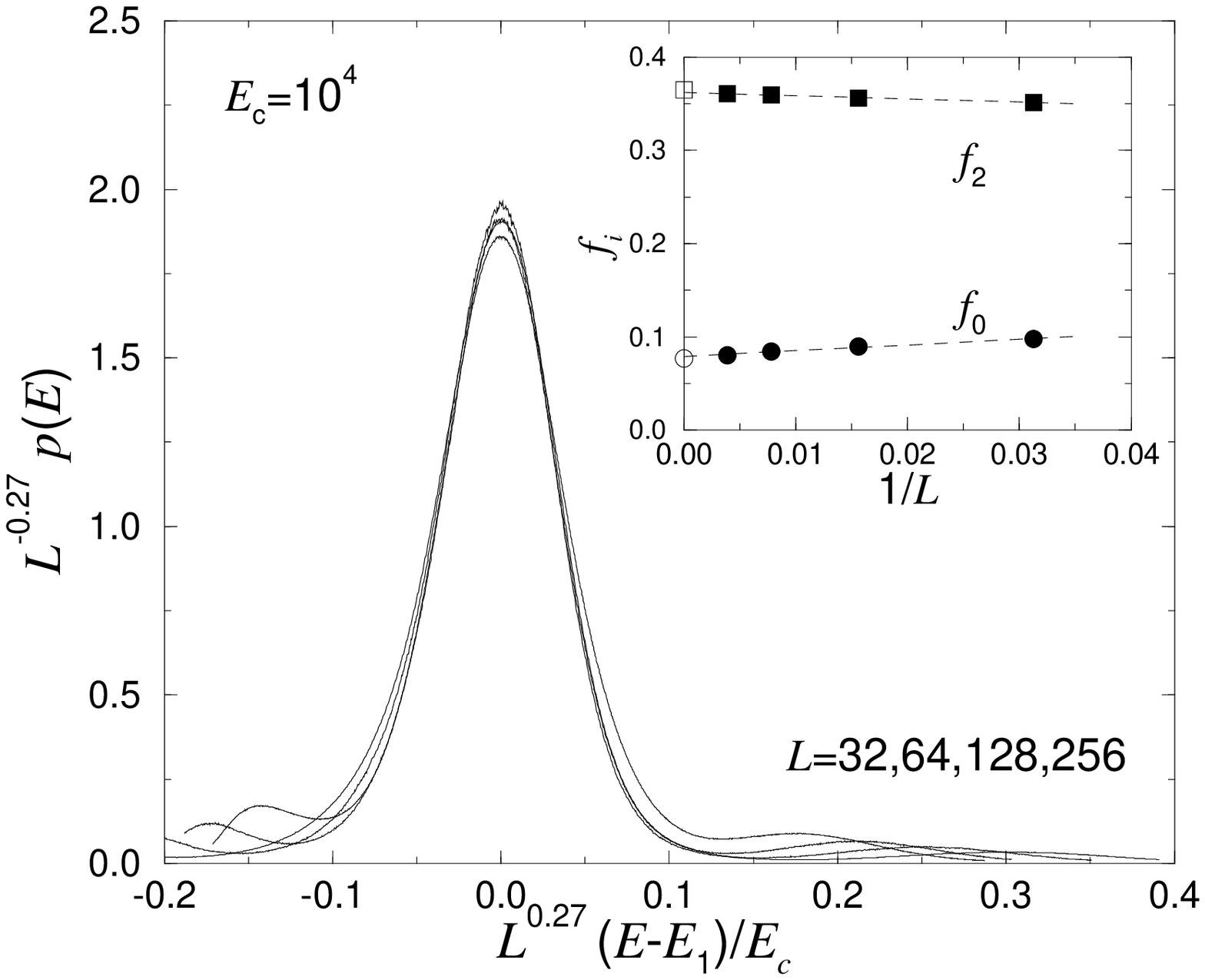}
 \caption{The finite-size scaling plot of the first maximum
	  of the energy distribution $p(E)$.
	  The inset displays the statistical weights~$f_0$ 
	  and $f_2$ as a 
	  function of the inverse system size.
	  The values of the infinite system (extrapolation
	  to the vertical axis) agree with those
	  of the Zhang model (open symbols) obtained 
	  from~\protect\cite{LUEB_3}.
 \label{fig:f_i_02}} 
\end{figure}

\begin{figure}[b]
 \epsfxsize=8.0cm
 \epsffile{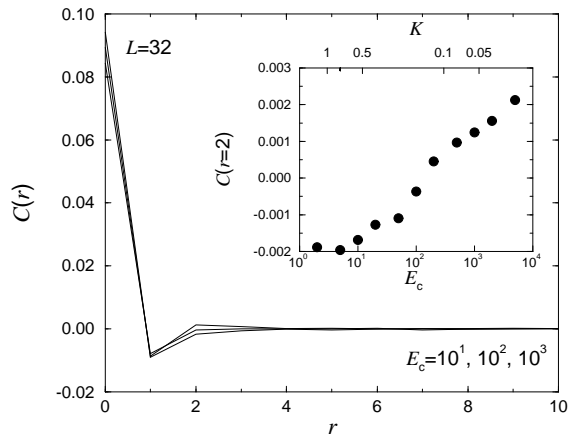}
 \caption{The correlation function $C(r)$ for different values
	  of the critical energy $E_{\rm c}$.
	  The inset displays the value $C(r=2)$ which is
	  an indicator for the crossover from the
          Manna scaling behavior [$C(r=2)<0$] 
	  to the Zhang scaling behavior [$C(r=2)>0$].
 \label{fig:h_corr_02}} 
\end{figure}

It is known from the Zhang model that the maxima of the
energy distribution~$p(E)$ increase with the system size
and that $p(E)$ is characterized by~$z$ $\delta$-peaks
for $L\to \infty$~\cite{LUEB_3}.
Therefore we measured the probability distribution~$p(E)$
for a large but fixed critical energy
($E_{\rm c}=10^4$) and for various system sizes.
In Fig.~\ref{fig:f_i_02} we present the 
rescaled distribution $L^y p(E)$ as a function
of $L^{-y} (E-E_i)$
around the first peak at $E_1=5/16$.
The resulting data collapse 
shows that similar to the pure Zhang model the peak
diverges for $L\to \infty$.
But in contrast to the pure Zhang model where the 
finite-size behavior of $p(E)$ is characterized by
an exponent $y\approx 0.6$~\cite{LUEB_3} we get 
for $E_{\rm c}=10^4$ the significant different 
value $y=0.27\pm 0.1$.
Similar results are obtained for the second and
third peak (not shown).
Further investigations have to show if the
finite-size scaling exponent~$y$ depends
on $E_{\rm c}$, i.e., if $y$ tends to $0.6$ for
$E_{\rm c}\to \infty$.

But nevertheless the finite-size scaling analysis
of the probability distribution yields that 
$p(E)$ is characterized by four $\delta$-peaks 
[Eq.~(\ref{eq:p_E_infty})]
and one can compare the statistical weight $f_i$ of each
peak with the values of the pure Zhang model.
Therefore we divided the interval $[0,E_{\rm c}]$
in four parts and measured in each part the 
area under the curve $p(E)$ for various system sizes.
The values of $f_0$ and $f_2$ are shown in the 
inset of Fig.~\ref{fig:f_i_02} as
a function of the inverse system size.
The same system size dependence was observed for
the pure Zhang model~\cite{LUEB_3}.
The extrapolation to $L\to \infty$ yields
the statistical weights $f_i$ and
the obtained values agree with those 
of the Zhang model (see vertical axis of the
inset of Fig.~\ref{fig:f_i_02}).

Finally we consider in this section the energy
correlations in the steady state.
The correlation function is defined as
\begin{equation}
C(\underline r) \; = \; 
\frac{\langle E_{\underline r'}\, E_{\underline r'+{\underline r}}\rangle
\, - \,
\langle E_{\underline r'}\rangle \langle E_{\underline r'+{\underline r}}\rangle}
{(E_{\rm c}-1)^2}.
\label{eq:correalation}
\end{equation}
In Fig.~\ref{fig:h_corr_02} we plot $C(r)$ along the
symmetry axis of the square lattice.
Starting from the autocorrelation peaks at $r=0$
the energies of neighboring sites ($r=1$) display an 
anti-correlated behavior.
This is caused by the toppling events
and is an typical property of sandpile models where the
energy of instable sites is reduced to zero.
In the inset of Fig.~\ref{fig:h_corr_02} we plot the
energy correlations for the distance $r=2$
as a function of $E_{\rm c}$ and $K$, respectively.
For small values of $E_{\rm c}$ we get negative correlations
which increases with the critical energy.
In the surrounding of $K^\star \approx 0.15$ the behavior
changes from an anti-correlated behavior ($C<0$) to
a correlated behavior ($C>0$).
Positive correlations between sites of the same sublattice
($r=2$) are a typical feature of sandpile models with an
isotropic energy transfer to the nearest neighbors.
Thus positive correlations within a sublattice 
indicate a Zhang like behavior 
whereas negative correlations are a
characteristic of the Manna model.
The analysis of the avalanche distributions in the 
next subsection 
reveals that the model exhibits the typical Manna scaling behavior
(for $E_{\rm c}=2$) as long as the energies are
anti-correlated, i.e., the scaling behavior changes
above the value $K^\star$.

\subsection{Avalanche distributions}

In the following we consider the avalanche 
distributions [Eq.~(\ref{eq:power_law})] and 
examine the behavior of the corresponding exponents~$\tau_x$ 
as a function of the critical energy.
Using the reported values of the 
exponents of the Zhang and Manna (for $E_{\rm c}=2$) 
model~\cite{MANNA_2,LUEB_3,CHESSA_2,LUEB_9}
we expect that the difference 
is nearly $\%1$ for the size exponent $\tau_s$,
less than $\%2$ for $\tau_a$, and nearly $\%5$
for the radius exponent $\tau_r$.
In the case of the duration exponent it seems
that both models are characterized by the 
value $\tau_t\approx 3/2$.
In the following we analysis the size and radius
exponents. 
The exponents are obtained from a regression analysis
of the corresponding probability distribution.

\begin{figure}[b]
 \epsfxsize=8.0cm
 \epsffile{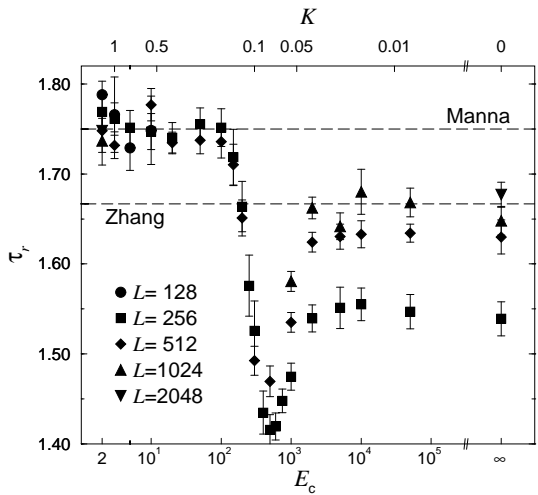}
 \caption{The avalanche exponent $\tau_r$ as a function of 
	  the critical energy $E_{\rm c}$ and the crossover
	  parameter~$K$, respectively.
	  The dashed lines indicate the value of the
	  Manna model for $E_{\rm c}=2$ (upper) and 
	  the Zhang (lower) model. 
	  The corresponding value of the pure Zhang model
	  is marked as $E_{\rm c}=\infty$.
 \label{fig:tau_r_Ec_02}} 
\end{figure}

In Fig.~\ref{fig:tau_r_Ec_02} we plot the exponent
$\tau_r$ for different values of the system size~$L$ 
and different critical energies~$E_{\rm c}$
(see also Table~\ref{table:tau_r}).
One can distinguish between three regimes:
for small values of the critical energy 
($E_{\rm c}\lesssim 100$) the exponents are independent
of $E_{\rm c}$ and display no significant
system size dependence.
This is the regime of the Manna universality class
and the obtained results agree with the observed universal behavior
for $E_{\rm c}\le 10$ in~\cite{LUEB_9}.

For large values of the critical energy 
($E_{\rm c}\gtrsim 5000$)
another regime occurs where the exponents are
nearly independent of $E_{\rm c}$ but display
a significant system size dependence.
These system size dependence agrees with the
observed $L$-dependence of the Zhang 
model [Eq.~(\ref{eq:zhang_tau_L})].
Thus we recover the third fingerprint
of the Zhang model for sufficiently large values 
of $E_{\rm c}$.
We call this regime in the following the Zhang regime.
Between the Manna and the Zhang regime we find
a third transient regime where the exponents
depend strongly on the critical energy
and on the system size.

\begin{figure}[t]
 \epsfxsize=8.0cm
 \epsffile{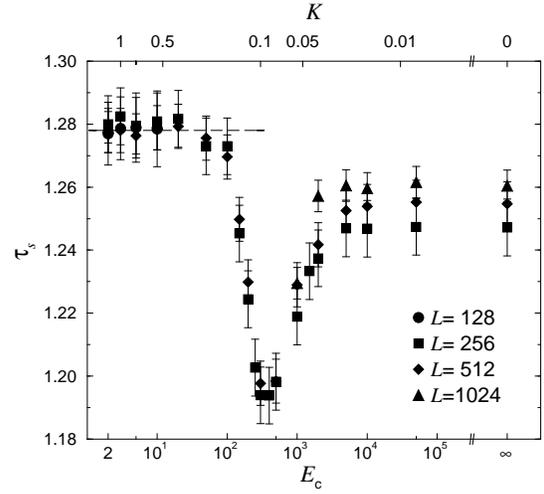}
 \caption{The avalanche exponent $\tau_s$ as a function of 
	  the critical energy $E_{\rm c}$ and the crossover
	  parameter~$K$, respectively.
	  The dashed line indicates the value of the
	  Manna model for $E_{\rm c}=2$. 
	  The corresponding values of the pure Zhang model
	  are marked as $E_{\rm c}=\infty$.
 \label{fig:tau_s_Ec_02}} 
\end{figure}

The avalanche size exponent~$\tau_s$ displays
a similar behavior (see Fig.~\ref{fig:tau_s_Ec_02}
and Table~\ref{table:tau_s}).
Again the Manna regime is characterized by independent
exponents.
Approaching a certain value ($E_{\rm c}\approx 100$)
the exponent depends strongly on $E_{\rm c}$ in the
transient regime.
In the third regime ($E_{\rm c}\gtrsim 5000$) we observe
the characteristic Zhang system size dependence.
Thus we see that the scaling behavior corresponds
to that of the Zhang model for large but finite 
values of the energy threshold $E_{\rm c}$. 
This means that adding stochasticity to the toppling 
rules of the Zhang model does not force a change of 
the universality class in all cases. 
Only a sufficiently large 
stochasticity is relevant, otherwise it is 
irrelevant and the scaling behavior is unchanged.

\begin{figure}[b]
 \epsfxsize=8.0cm
 \epsffile{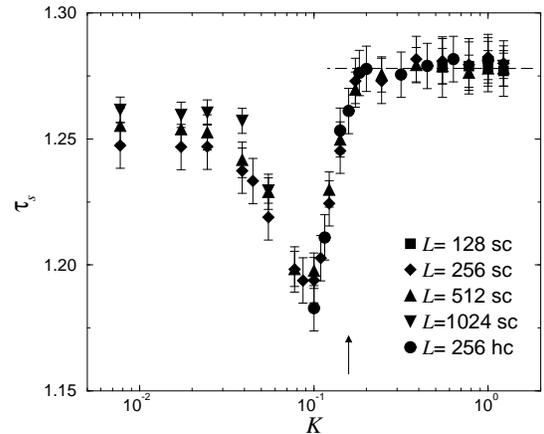}
 \caption{The avalanche exponent $\tau_s$ as a function of 
	  the scaling parameter $K$ for the 
	  simple cubic (sc) and honeycomb (hc) lattice.
	  The dashed line indicates the value of the Manna
	  model for $E_{\rm c}=2$ and the arrow marks the 
	  value $K^\star=0.15$ obtained from the analysis 
	  of the correlation function 
	  [Eq.~(\protect\ref{eq:correalation})].
 \label{fig:tau_s_K_01}} 
\end{figure}

Considering the behavior of the exponents
$\tau_s$ and $\tau_r$ in the transient regime it
seems that the exponents increase with the system size.
Especially the "over-shooting" effect decreases
with increasing~$L$ and usually one would expect that
this regime disappears for $L\to\infty$.
Unfortunately it is impossible to conclude from 
the numerical results
whether the transient regime still exist in the 
thermodynamic limit.
Further investigations are needed to answer this
question seriously.

In the last figure we plot the size exponent $\tau_s$
for two different lattice types as a function of the
crossover parameter $K$.
As one can see both curves display the same behavior, i.e.,
the transition from the Manna regime to the transient
regime takes place at $K^\star \approx 0.2$ independent
of the lattice type.
This value is in agreement with the value $K^\star\approx 0.15$
obtained from the analysis of the correlation 
function,
i.e., the change of the scaling behavior coincides
with the change from negative to positive correlations $C(r=2)$.
Thus we conclude that the Manna scaling behavior is 
strongly connected to the anti-correlations between 
next nearest neighbors.

\begin{table}[t]
\caption{Some values of the avalanche radius exponent $\tau_{r}$ for 
various values of the critical energy $E_{\rm c}$ and different
sizes~$L$ of the square lattice.
The corresponding value of the pure Zhang model
is marked as $E_{\rm c}=\infty$.}
\label{table:tau_r}
\begin{tabular}{llll}
 $E_{\rm c}$	& $L=256$	&	$L=512$   &	$L=1024$  \\
\tableline
2	& 1.769	& 1.748	& 1.737	\\
5	& 1.752	& 1.732	& 	\\
10	& 1.747	& 1.777	& 	\\
20	& 1.740	& 1.735	& 	\\
50	& 1.755	& 1.738	& 	\\
100	& 1.752	& 1.736	& 	\\
200	& 1.664	& 1.651	& 	\\
500	& 1.416	& 1.470	& 	\\
1000	& 1.475	& 1.535	& 1.581	\\
2000	& 1.539	& 1.624	& 1.662	\\
5000	& 1.551	& 1.630	& 1.642	\\
10000	& 1.555	& 1.633	& 1.680	\\
50000	& 1.547	& 1.634	& 1.668	\\
$\infty$& 1.539	& 1.630	& 1.648	\\
\end{tabular}
\end{table}

\begin{table}[b]
\caption{Some values of the avalanche size exponent $\tau_{s}$ for 
various values of the critical energy $E_{\rm c}$ and different
sizes~$L$ of the square lattice.
The corresponding value of the pure Zhang model
is marked as $E_{\rm c}=\infty$.}
\label{table:tau_s}
\begin{tabular}{llll}
 $E_{\rm c}$	& $L=256$	&	$L=512$   &	$L=1024$  \\
\tableline
2	& 1.280	& 1.278	& 1.279	\\
5	& 1.279	& 1.276	& 	\\
10	& 1.281	& 1.279	& 	\\
20	& 1.282	& 1.279	& 	\\
50	& 1.273	& 1.276	& 	\\
100	& 1.273	& 1.270	& 	\\
200	& 1.224	& 1.230	& 	\\
500	& 1.198	& 1.198	& 	\\
1000	& 1.219	& 1.229	& 1.230	\\
2000	& 1.237	& 1.242	& 1.257	\\
5000	& 1.247	& 1.253	& 1.261	\\
10000	& 1.246	& 1.254	& 1.260	\\
50000	& 1.248	& 1.255	& 1.262	\\
$\infty$& 1.247	& 1.255	& 1.260	\\
\end{tabular}
\end{table}

\section{Summary}
\label{summary}

We investigated the scaling behavior of the Manna
model as a function of the threshold condition.
Increasing the value of the critical energy
a crossover takes place from the  universality
class where the energy transfer is undirected
on average to the universality class of an
isotropic energy transfer (Zhang model).
For sufficiently large thresholds all characteristic
of the Zhang model could be recovered, namely the
quantization of the energy values in the steady 
state, the statistical weights of the energy 
quantums in the thermodynamic limit and the
observed system size dependence of the avalanche 
exponents.
Our analysis suggests that the scaling behavior of the stochastic 
sandpile model is connected to the negative 
correlations between next nearest lattice sites.

I would like to thank A.~Hucht for useful discussions.

\end{document}